\documentclass[aps,pre,twocolumn,,showpacs,showkeys,floatfix]{revtex4}%

\usepackage{epsfig}
\usepackage{times}
\usepackage{amssymb}
\usepackage{colordvi}
\usepackage{color}     
\usepackage{amsbsy}
\usepackage{amsmath}
\usepackage{latexsym}
\usepackage{graphicx}

\newcommand{\preprintonly}[1]{}

\newcommand{\RESTORECOLOR}{}   

\newcommand{\FINALREF}[1]{\Blue{\ref{#1}}\RESTORECOLOR}

\newcommand{\SKIPBIBITEM}[2]{}

{\rm \addcolor $\bullet|$ \blackcolor}                

\newcommand{\HIDDEN}[1]{}
                          {]\end{quotation}\blackcolor\renewcommand{\RESTORECOLOR}{}\rm} 

\newcommand{\stl}[1]{{\hspace{0.2em}
\stackrel{ \setbox0=\hbox{\hspace{0.06em}$\displaystyle
#1$\hspace{0.06em}} \setbox1=\hbox{\vrule width\wd0 height0.08ex depth0pt} \vrule width0.08ex
height0.08ex depth0.475ex \box1 \vrule width0.08ex height0.08ex
depth0.475ex } {#1}\hspace{0.2em}}}

\newcommand{\JOINED}[1]{}

 \newcommand{\X}{{\bf X}}
 \newcommand{\ave}[1]{\left\langle #1\right\rangle}
 \newcommand{\avee}[1]{\langle #1\rangle}
 
 \newcommand{\bea}{\begin{eqnarray}}
 \newcommand{\eea}{\end{eqnarray}}
 \newcommand{\be}{\begin{equation}}
 \newcommand{\ee}{\end{equation}}
 
 \newcommand{\F}{f^\textrm{GM}}

 \newcommand{\dphi}{\delta\varphi}

 \renewcommand{\k}{\textbf{k}}
 
 \newcommand{\mbf}[1]{\mbox{\boldmath{$#1$}}}
 
 \newcommand{\q}{\tilde{\bf q}}
 \newcommand{\x}{\textbf{x}}
 \newcommand{\ONE}{\mathbf{I}}

 \newcommand{\e}{{\bf e}_\parallel}

 \newcommand{\gammapara}{\cpa}

 \newcommand{\beas}[1]{\begin{subequations}\label{#1}\bea}
 \newcommand{\eeas}{\eea\end{subequations}}
 
 \newcommand{\gammaperpnew}{\cpe}

 \renewcommand{\c}{{\bf c}}
 \newcommand{\cc}{c}
 
 \renewcommand{\q}{{\bf q}}

 \renewcommand{\u}{\textbf{u}}

 \newcommand{\E}{{\bf e}_\perp}

 \newcommand{\aveeGMS}[1]{\avee{#1}}
 
 \renewcommand{\Re}{{\rm Re}}
 \renewcommand{\Im}{{\rm Im}}

 \renewcommand{\parallel}{{\mbox{\tiny$\|$}}}
 \renewcommand{\perp}{{\mbox{\tiny$\bot$}}}

 \newcommand{\stress}{\mbf{\sigma}}

 \newcommand{\plus}{\oplus}
 \newcommand{\minus}{\ominus}

 \newcommand{\HIDE}[1]{}

 \renewcommand{\bar}[1]{\stl{#1}}
 \renewcommand{\overline}[1]{\stl{#1}}
 
 \newcommand{\cphi}{c_\perp}
 \newcommand{\cpe}{c_\phi}
 \newcommand{\cpa}{c_\parallel}
 
 \newcommand{\irrINSERT}{Adopting the notation in
 \cite{tensorrechnung2,mkbook} the distribution function is written as
 a sum over $n$--fold contracted products of $n$th rank tensors, 
 $f(\c)=f_0(c)\sum_{k,n=0}^\infty \ave{\mbf{\phi}_k^n}\odot^n\mbf{\phi}_k^n(\c)$
 with $\ave{\mbf{\phi}_k^n}=\int f(\c)\mbf{\phi}_k^n\,d^3c$ and base functions
 $\mbf{\phi}_k^n(\c)=l_k^n L_k^{n+1/2}(c^2)\stl{\otimes^n\c}$, 
 where $L_k^n$ are the associated Laguerre ($k$th order) polynomials \cite{abramowitz}, 
 $\otimes^n\c$ denotes the $n$--fold tensor product, and 
 $\stl{\textbf{a}}$ denotes the irreducible part of a tensor $\textbf{a}$. 
 For the explicit construction of $n$th rank irreducible tensors $\stl{\otimes^n\c}$
 see page 160 of \cite{mkbook}.
 The normalization coefficients evaluate as $l_k^n=(\sqrt{\pi}k!(1+2n)!!/[2(k+n+1/2)!n!])^{1/2}$. 
 The base function $\mbf{\phi}_k^n(\c)$ is thus a $(2k+n)$th order polynomial in $c$.
 The lowest order base functions read $\mbf{\phi}_0^0 =1$, $\mbf{\phi}_0^1 =\sqrt{2}\c$, 
 $\mbf{\phi}_1^0 =\sqrt{2/3}(3/2-c^2)$, $\mbf{\phi}_1^1 =(2/\sqrt{5})(5/2-c^2)\c$, 
 and $\mbf{\phi}_0^2 =\sqrt{2}\stl{\c\c}$. Density, velocity, temperature, heat flux, and stress tensor 
 are related to the moments as follows: $\tilde{n}=\ave{\mbf{\phi}_0^0}$, 
 $\tilde{\u}=\ave{\mbf{\phi}_0^1}/\sqrt{2}$, 
 $\tilde{T}=\ave{\mbf{\phi}_1^0}\sqrt{3/2}$, 
 $\q=\ave{\mbf{\phi}_1^1}$, and $\mbf{\sigma}=\ave{\mbf{\phi}_0^2}/\sqrt{2}$. 
 The distribution function is then split into (orthogonal) parts as 
 $f(\c)=f^\textrm{LM}(\c)+\delta f^\textrm{Grad}(\c)+\delta f^\textrm{rest}(\c)$
 with $f^\textrm{LM}(\c)\equiv f_0(c)(\ave{\mbf{\phi}_0^0}\mbf{\phi}_0^0 + \ave{\mbf{\phi}_0^1}\mbf{\phi}_0^1 + \ave{\mbf{\phi}_1^0}\mbf{\phi}_1^0)$ and $\delta f^\textrm{Grad}(\c) \equiv 
 f_0(c)(\ave{\mbf{\phi}_1^1}\mbf{\phi}_1^1 + \ave{\mbf{\phi}_0^2}\mbf{\phi}_0^2)$,
 while the sum in $\delta f^\textrm{rest}(\c)=\sum_{k,n} \ave{\mbf{\phi}_k^n}\odot^n \mbf{\phi}_k^n(\c)$
 extends over the remaining $(k,n)$--pairs. Density, velocity, and temperature
 are therefore determined by $f^\textrm{LM}$ alone, and $\delta f$
 automatically obeys constrains such as orthogonality requirement
 $\int \delta f(\c) \mbf{\phi}_1^0\,d^3c=0$ and also $\int \delta f(\c)\xi(\c)d^3c=0$, 
 as mentioned in the text part. These conditions become redundant
 ones calculations are performed using the particular basis $\mbf{\phi}_k^n$.
 For Maxwell molecules, the dependence on the polar angle $\phi$ can be 
 included by replacing $P_l(z)$ by $e^{im\phi} P_l^m(z)$ involving the associated Legendre
 polynomials \cite{abramowitz}, and the
 eigenvalues are independent of $m$. Then, these base function reduce to the 
 eigenfunctions $\Psi_{r,l}(c,z)$ (\ref{eigenfunctions}) of the Maxwell gas
 }

 \newcommand{\calc}{The integrals listed
 in Tab.\ \ref{tab1} obey the following decoupling rules: 
 \bea \int(\cpa^2-\frac{1}{3}c^2) \delta X_n d^3c &\propto&1-\delta_{n,4}, \nonumber \\
 \int \cpa \cphi \delta X_n d^3c&\propto&\delta_{n,4},\nonumber \\
 \int \cpa(c^2-\frac{5}{2})\delta X_n d^3c&\propto&1-\delta_{n,4},\nonumber \\
 \int \cphi(c^2-\frac{5}{2})\delta X_n d^3c&\propto& \delta_{n,4}.
 \eea}

\begin{document}

\title{Boltzmann equation and hydrodynamic fluctuations}
\date{ {\small 2009-07-07 14:54:44} 
} 

\author{M. Colangeli}   
\affiliation{Polymer Physics, Department of Materials, ETH Z\"urich, CH-8093 Z\"urich, Switzerland}

\author{M. Kr\"oger} \email{mk@mat.ethz.ch} \homepage{http://www.mat.ethz.ch} 
\affiliation{Polymer Physics, Department of Materials, ETH Z\"urich, CH-8093 Z\"urich, Switzerland}

\author{H.C. \"Ottinger}   
\affiliation{Polymer Physics, Department of Materials, ETH Z\"urich, CH-8093 Z\"urich, Switzerland}


\newcommand{\INLINEFIGURE}[1]{#1}
\begin{abstract}
We apply the method of invariant manifolds to derive equations of generalized hydrodynamics from the linearized Boltzmann equation and determine exact transport coefficients, obeying Green-Kubo formulas. Numerical calculations are performed in the special case of Maxwell molecules. We investigate, through the comparison with experimental data and former approaches, the spectrum of density fluctuations and address the regime of finite Knudsen numbers and finite frequencies hydrodynamics.

\end{abstract}


\maketitle

 \section{Introduction} \label{secI}

 The Boltzmann equation (BE) lies at the basis of classical and
 quantum kinetic theory of gases. It provides a detailed picture of the time
 evolution of a dilute gas towards a thermal equilibrium state,
 which constitutes the essence of the H-theorem. This celebrated
 result gave rise, historically, to the first clear insurgence of
 irreversibility into deterministic equations of motion. The
 nonlinear integro-differential nature of the BE prevented, so far, an exact
 solution. Perturbative methods and kinetic toy models have
 been devised such to get partial answers. The
 Chapman-Enskog expansion (CE) was, in particular, the first
 important success in this direction \cite{chapman}, as it allowed
 to consistently derive hydrodynamics laws from their microscopic
 counterpart and to obtain rigorous expressions for transport
 coefficients. The CE method is based upon a perturbative expansion
 of the distribution function in terms of the Knudsen number
 $\varepsilon$, defined as the ratio between the mean free-path and a
 macroscopic hydrodynamic length. This is supposed to be a
 ``smallness'' parameter, in that the series converges only for
 $\varepsilon \rightarrow 0$. By increasing the order of the
 expansion one should not expect to capture larger extents of the
 ``true'' solution of the BE, since, as it was pointed out by Bobylev
 \cite{Bobylev82}, one has to face divergencies of the acoustic modes
 in the dispersion relation, which are inherently related to the
 procedure of truncation. In order to tackle this unphysical feature
 of post-Navier Stokes hydrodynamics, some regularization methods
 were borrowed from functional analysis in order to restore the
 H-Theorem \cite{Bobylev06}. Another route, which attempts a non--perturbative 
 approach to solve the BE, is based upon the notion
 of Invariant Manifold \cite{karlinbook}. Through this method, one
 assumes \emph{a priori} a separation of the hydrodynamic time scale
 and the kinetic time scale and postulates the existence of a stable
 Invariant Manifold (IM) in the space of distribution functions,
 which is parameterized with the values of the hydrodynamic fields:
 particle number, velocity, and temperature. In this paper we address the
 study of the spectrum of hydrodynamic excitations in a Maxwell gas,
 employing the latter non--perturbative approach, which will allow to
 find exact transport coefficients at arbitrary length scales. The
 paper is organized as follows: in Sec. \FINALREF{secII} we review the
 eigenvalue problem associated with the linearized Boltzmann equation
 and recall that hydrodynamic modes at finite wavevector can be obtained
 as eigenvalues of a perturbed linear operator. Next, in Sec.\ \FINALREF{secIII}, we motivate and derive 
 the invariance equations (details collected in Appendix \FINALREF{appendixA}) and 
 consider the case of Maxwell molecules (Sec.\ \FINALREF{solving}), 
 whose associated eigenvalue problem for the unperturbed operator is analytically
 solvable (Appendix \FINALREF{appendixB}). Postulating the existence of an IM, we solve the
 eigenvalue problem for arbitrary wavevectors and find generalized transport coefficients which recover the Green-Kubo
 formulas in Sec.~\FINALREF{moresolving}. Further, in Section \FINALREF{secIV}, we determine the spectrum of density fluctuations and formulate a hypothesis about the features of finite wavelengths hydrodynamics. Conclusions are drawn in Sec. \FINALREF{secV}.

 \section{Eigenvalue problems for the Boltzmann equation and hydrodynamics} \label{secII}
 
 The dynamics of the fluctuations of hydrodynamic fields (particle number, momentum, temperature) as induced by the properties of the underlying microscopic or kinetic equation, is an important issue in statistical mechanics which dates back to the seminal work by Onsager \cite{Onsager}.
 In this section we focus upon the BE and show in a general setting how it features equilibration through some generalized frequencies (inverse of characteristic collision times). The way how these generalized frequencies give rise and affect the decay rates of some collective fluctuations (hydrodynamic modes) of the macroscopic fields, is still an issue which lacks a rigorous foundation.
 The reason is that the hydrodynamic equations, as derived from the BE, are not closed and hence, some (semi-phenomenological) approximations for higher order moments need to be included. In particular, the celebrated Navier-Stokes-Fourier (NSF) approximation was the first historically relevant attempt in this direction.
 We start by reviewing some results formerly obtained by Resibois \cite{resibois}. He shed some preliminary light upon the connection between the generalized frequencies and the hydrodynamic modes by solving, via perturbation theory, the eigenvalue problems associated \emph{independently} to the BE and to the NSF equations of hydrodynamics.
 
 \subsection{The Boltzmann equation} \label{2A}
 
 The BE reads:
 \be \partial_{t} f = -\textbf{v}\cdot\nabla f+ Q[f,f], \label{BE} \ee
 where $Q$ denotes a non-linear integral  collision operator. We
 introduce the thermal velocity $v_{T}=\sqrt{2 k_{B}T_{0}/m}$, the
 dimensionless peculiar velocity
 $\textbf{c}=(\textbf{v}-\textbf{u}_{0})/v_{T}$ and the equilibrium
 values of macroscopic fields: equilibrium particle number $n_{0}$,
 equilibrium mean velocity $\textbf{u}_{0}=\textbf{0}$, and equilibrium
 temperature $T_{0}$. The global Maxwellian is defined as:
 $f^\textrm{GM}=(n_{0}/v_{T}^{3})f_{0}(c)$ where
 $f_{0}(c)=\pi^{-3/2}e^{-c^{2}}$ denotes a Gaussian in
 velocity space ($c\equiv|\textbf{c}|$). We consider only small
 disturbances from the global equilibrium. After passing over to Fourier space, we write the
 distribution function (cf. also Tab.\ \FINALREF{tab:notation}) as:
 \be f(\textbf{k},\textbf{c},t)=f^\textrm{LM}+\delta f, \label{linear} \ee
 where $f^\textrm{LM}$ denotes the local Maxwellian to be made precise in Sec.~\FINALREF{secIII}, 
 and $\delta f$ the
 deviation from local equilibrium. An alternative notation is introduced 
 via $\delta f = f^\textrm{GM}\delta \varphi$.
 Considering a co-moving reference frame and linearizing the collision operator around global equilibrium, one obtains from (\FINALREF{BE})
 \be \frac{1}{v_T} \partial_{t}  f = -i \textbf{k} \cdot \textbf{c}f+
 \hat{L}\delta f, \qquad \hat{L}=\frac{1}{v_{T}} L, \label{BElin} \ee
 where we made use of the fact that $Lf^\textrm{LM}=0$. 
 The linearized Boltzmann collision operator, $L$, assumes the form
 \bea L\delta f&=& \int\int d\Omega d \textbf{c}_{1} \sigma(\Omega,g)
 g f^\textrm{GM}(c_{1}) \times \nonumber \\
 && [\dphi(\textbf{k},\textbf{c})+\dphi(\textbf{k},\textbf{c}_1) - \dphi(\textbf{k},\textbf{c}')-\dphi(\textbf{k},\textbf{c}_{1}'].
 \label{linBoltz} \eea
 Here, $\sigma(\Omega,g)$ is the scattering cross section, $g\equiv |v-v_1|$, and $v$, $v_1$ are the velocities of the
 particles entering the binary collision. In the remainder ot this section, we will focus our attention upon the operator
 $\mathrm{\Lambda}\equiv \hat{L}-i\textbf{k}\cdot\textbf{c}$, whose spectral properties determine the time evolution of the distribution function. This is readily seen by considering the Laplace time transform of Eq.\ \FINALREF{BElin} (to be further 
 discussed in Sec.\ \FINALREF{secIV}) and by inspection of the inverse transform, which reads as:
 \be f(\textbf{k},\textbf{c},t)=\frac{1}{2\pi i}\oint 
 \frac{e^{z t}}{(z-\mathrm{\Lambda})}dz f(\textbf{k},\textbf{c},0), \label{sol} \\
 \ee
 where the closed path encircles all the poles of the integrand function.      
 Through the Spectral Theorem, we regard these poles as coinciding with the spectrum of the operator
 $\mathrm{\Lambda}$. The investigation of the spectral properties of such an
 operator is, in fact, a longstanding issue in kinetic theory \cite{Blatt}.
 In order to study the eigenvalue problem associated with $\Lambda$, we introduce
 the Fourier time transform of the distribution function 
 $f(\textbf{k},\textbf{c},\omega)=\int_{-\infty}^{\infty}e^{-\omega t} f(\textbf{k},\textbf{c},t)dt$,
 where $\omega$ defines a complex valued quantity. 
 Then, (\FINALREF{BElin}) reduces to:
 \be \Lambda f = \omega f \label{start}, \ee
 which constitutes the starting point of our analysis.
   
 In the present paper, functions $f_{\mu}=f^\textrm{GM}\varphi_{\mu}$ 
 will be regarded as vectors in a Hilbert space, whose scalar product is defined by:
 \be \langle f_{1}|f_{2}\rangle=\frac{1}{n_{0}}\int
 (f^\textrm{GM})^{-1}f_{1}(\textbf{c})f_{2}(\textbf{c})d^{3}
 v. \label{prod} \ee
 The spectrum of $\Lambda$ is analytic in $k=0$ and it can
 be shown to contain a $D+2$--fold degeneracy at the origin,
 corresponding to local conserved quantities. In order to solve the
 eigenvalue problem associated with (\FINALREF{start}), it is worth first
 to attempt the analysis of the long-wavelength limit
 $k\rightarrow0$: \be
 \hat{L}\Psi_{i}(\textbf{c})=\lambda_{i}\Psi_{i}(\textbf{c}). \label{eigen} 
 \ee The operator $\hat{L}$ is found to be symmetric and negative
 semidefinite with respect to the scalar product (\FINALREF{prod}), hence eigenfunctions are orthogonal and form a complete set. In
 particular, a subset of them, which spans a $(D+2)$--dimensional
 subspace of the Hilbert space can be found corresponding to the
 degenerate zero eigenvalue. These are the collision invariants
 $f^\textrm{GM}\X^0$, with $\X^0$ denoting a set of lower order Sonine (or associated Laguerre) polynomials:
 \be\X^0=\left[1,2\textbf{c},\left(c^{2}-\frac{3}{2}\right)\right]\label{X0} \ee
 (see also Eq.\ \FINALREF{defX0}).
 A perturbative approach is followed in order to extract those eigenvalues, denoted hereafter by $\omega_\textrm{hydro}$,
 which reduce to zero in the long wavelength limit, from the full spectrum of $\Lambda$. The yet unknown eigenfunctions and eigenvalues are expanded in powers of the wavevector $\textbf{k}$:
 \bea
 |\Psi_{\alpha}\rangle&=&|\Psi_{\alpha}^{(0)}\rangle+k|\Psi_{\alpha}^{(1)}\rangle+k^{2}|\Psi_{\alpha}^{(2)}\rangle+\dots, \nonumber \\
  \omega_{\alpha}&=&\omega_{\alpha}^{(0)}+k \omega_{\alpha}^{(1)}+k^{2}\omega_{\alpha}^{(2)}+\dots,
 \label{expand} \eea
 where $|\Psi_{\alpha}^{(0)}\rangle$ denotes a linear combination of
 the eigenfunctions of the unperturbed system. The
 result of this standard procedure is a polynomial expression for the
 set $\{\omega_{\alpha}\}$ of hydrodynamic modes, up to second order:
 \bea
 \omega_{1}&=&i c_{0}k-k^{2}\langle\Psi_{1}^{(0)}|(c_{x}-c_{0})\frac{1}{\hat{L}}(c_{x}-c_{0})|\Psi_{1}^{(0)}\rangle, \nonumber \\
 \omega_{2}&=&- i c_{0}k-k^{2}\langle\Psi_{2}^{(0)}|(c_{x}+c_{0})\frac{1}{\hat{L}}(c_{x}+c_{0})|\Psi_{2}^{(0)}\rangle, \nonumber \\
 \omega_{3}&=&-k^{2}\langle\Psi_{3}^{(0)}|c_{x}\frac{1}{\hat{L}}c_{x}|\Psi_{3}^{(0)}\rangle, \nonumber \\
 \omega_{4}&=&-k^{2}\langle\Psi_{4}^{(0)}|c_{x}\frac{1}{\hat{L}}c_{x}|\Psi_{4}^{(0)}\rangle, \nonumber \\
 \omega_{5}&=&-k^{2}\langle\Psi_{5}^{(0)}|c_{x}\frac{1}{\hat{L}}c_{x}|\Psi_{5}^{(0)}\rangle,
 \label{expand2} \eea
 where $c_{0}=(5 k_{B}T_{0}/3 m)^{\frac{1}{2}}$ is the speed of sound of an ideal gas.

 \begin{table}[bth]
 \begin{tabular}{ccccccccccc}
 \hline\hline
   $f$ & = & \multicolumn{3}{c}{$f^\textrm{LM}$}  & + &  \multicolumn{1}{c}{$\delta f$}&
   \\[-1mm]
       &   & \multicolumn{3}{c}{$\overbrace{\hspace*{3.5cm}}$} \\[-1mm]
       & = & $f^\textrm{GM}$ & + & $f^\textrm{GM}\varphi_0$  & + & \multicolumn{1}{c}{$f^\textrm{GM}\delta \varphi$} & \\
      
       & = & $f^\textrm{GM}$ & $+ $ &   $f^\textrm{GM}\X^0\cdot\x$ &  + & $f^\textrm{GM}\delta \X\cdot\x$ &
       \\[-3mm]
       &   &   &  & \multicolumn{3}{c}{$\underbrace{\hspace*{4.8cm}}$}
       \\
       & = & $f^\textrm{GM}$ & $+$ & \multicolumn{3}{c}{$f^\textrm{GM} \triangle \X\cdot\x$} & \\
       & = & $f^\textrm{GM}$ & $+$ &  & $\triangle f$ & \\
 \hline\hline
 \end{tabular}
 \caption{Notation used in this manuscript. Terms have been grouped and abbreviated 
 as depicted in this table. $f^\textrm{GM}$ and $f^\textrm{LM}$ denote global and local maxwellian, 
 respectively, and $\triangle f$ and $\delta f$ their ``distance'' from $f$. The third row informs about
 the closure discussed in this manuscript, while $\x$ is a set of lower order moments of $f$.}
 \label{tab:notation}
 \end{table}

 \subsection{Linear Hydrodynamics}

 We denote by $[n_{k},\mathbf{u}_{k},T_{k}]$ the Fourier
 transforms of the hydrodynamics fields, for instance:
 $n_{k}=\int_{-\infty}^{\infty}dt\int_{-\infty}^{+\infty}d \mathbf{r}
 e^{-\omega t-i \mathbf{k}\cdot\mathbf{r}}\delta n(\mathbf{r},t)$,
 where $\delta n(\mathbf{r},t)$ is the fluctuation at time $t$ and
 point $\mathbf{r}$ of the local particle number. The equations of
 hydrodynamics considered in \cite{resibois}, are the linearized Navier-Stokes-Fourier (NSF) equations, which represent balance equations for particle number density, momentum and kinetic energy endowed with specific constitutive equations for the stress tensor and heat flux:
 
 \beas{linhyd}
 \omega n_{k}(\omega)&=&- i n_{0}\mathbf{k}\cdot\mathbf{u}_{k}(\omega),\\
 \omega \mathbf{u}_{k}(\omega)&=& 
 -i\frac{\mathbf{k}}{n_{0}}\left(\frac{\partial P}{\partial n}\right)_{T}n_{k}(\omega)-i\frac{\mathbf{k}}{n_{0}}
 \left(\frac{\partial P}{\partial T}\right)_{n}T_{k}(\omega)\nonumber\\
 &&
 -\frac{\eta}{n_{0}} k^{2}\mathbf{u}_{k}(\omega)
 -\frac{\mathbf{k}}{n_{0}}\left(\zeta+\frac{1}{3}\eta\right)\mathbf{k}\cdot\mathbf{u}_{k}(\omega), \\
 \omega T_{k}(\omega)&=& 
 -i\frac{1}{n_{0}}\frac{T_{0}}{C_{v}}\mathbf{k}\cdot\mathbf{u}_{k}(\omega)
 -\frac{\kappa}{C_{v}n_{0}}k^{2}T_{k}(\omega),
 \eeas
 where $\zeta$ and $\eta$ are
 respectively the bulk and shear viscosity, $C_{v}$ is the specific
 heat at constant volume and $\kappa$ is the thermal conductivity.
 Solving the Eqs.\ (\FINALREF{linhyd}) amounts to
 determine the eigenvalues of a $5\times 5$ non Hermitian matrix, which
 represent the decay rates of the collective excitations. The
 intuition enlightened in the paper \cite{resibois} was to put into
 correspondence the macroscopic eigenvalues with their microscopic
 counterpart, obtained from (\FINALREF{eigen}) by application of
 perturbation theory. This identification allowed to find an
 approximate expression for transport coefficients only in terms of
 the one-body distribution function which turned out to be equivalent
 to reduced expressions determined by many-body autocorrelation
 functions. These coefficients properly recover the Chapman-Enskog
 expressions from classical kinetic theory.
 Within the above construction it is found that the decay rates of
 hydrodynamic modes in the NSF approximation are quadratic in the wave
 vector Re$(\omega)\propto -k^{2}$ and unbounded. The use of a suitable projector, on the other hand, outlined
 in the next section, allows us to find proper asymptotics and paves the way to
 solve the eigenvalue equation (\FINALREF{start}) as well as to determine
 exact transport coefficients.
 
 \section{The Invariant Manifold technique} \label{secIII}
 The notion of invariant manifold is a generalization of normal
 solution in the Hilbert and Chapman-Enskog method. Given a
 dynamical system
 \be \frac{df}{dt}=J(f),\label{dynsys} \ee
 where $J(f)$ is the vector field which induces the motion in the
 space of distribution functions $\textit{U}$.
 Given bounded and smooth functions $\x(\textbf{r},t)$ 
 we define the \textit{locally finite-dimensional} manifold
 $\Omega\subset\textit{U}$ as the set of functions
 $f(\x(\textbf{r},t),\textbf{c})$. Hence, we will only consider sets of
 distribution functions whose dependence upon the space variable
 $\textbf{r}$ is parameterized through some ``moments''
 $\x(\textbf{r},t)$. As it will be discussed in Sec.~\FINALREF{comparison}, once we identify such coarse-grained fields $\x(\textbf{r},t)$ with the hydrodynamic fields, postulating their existence corresponds to invoking the hypothesis of local thermodynamic equilibrium. Hence, the extent of our predictions is inherently restricted to lenght scales wherein the concept of a field as ensemble average over a statistically significant number of particles is still meaningful. Let us denote by $T_{w}$ the tangent space to the manifold $\Omega$ at the point $w$ of the phase space, and let us introduce a projection operator $P$ which, when acting on $J(f)$, describes the motion of the vector field along the manifold. The dynamics is, hence, splitted into a fast motion on the affine subspace $w+\textrm{ker}[P]$ and a slow motion, which occurs along the tangent space $T_{w}$ \cite{karlinbook}. 
 The set of eigenvalues $\omega_\textrm{hydro}$ is determined as follows:
 \begin{enumerate}
   \item We seek for an invariant manifold $\Omega\subset\textit{U}$ such
 that the following Invariance Equation (IE) is fulfilled:
 \be (\textbf{1}-P)\Lambda \Delta f=0 \label{IE} \ee
 
 where $\triangle f\equiv f-f^\textrm{GM}$ (cf. also Tab.\ \FINALREF{tab:notation}).
   \item After determining the nonequilibrium distribution function from Eq. (\FINALREF{IE}), we derive equations of linear
   hydrodynamics via integration of the kinetic equation (\FINALREF{BElin}). By construction, the decay rates of the macroscopic
  excitations then coincide with $\omega_\textrm{hydro}$.
 \end{enumerate}
 
 Let  $\x=[\tilde{n},\tilde{\textbf{u}},\tilde{T}]$ denote the set of dimensionless hydrodynamic fluctuations:
 $\tilde{n}\equiv(n-n_{0})/n_{0}=$ (particle number
 perturbation), $\tilde{\textbf{u}}\equiv \textbf{u}/v_{T}=$ (velocity perturbation) and
 $\tilde{T}\equiv(T-T_{0})/T_{0}$ (temperature perturbation). 
 Further, we split the mean velocity $\tilde{\u}$ uniquely as $\tilde{\u}=u^\parallel\e+u^\perp\E$, where the unit vector $\e$ is parallel to $\k$, and $\E$ orthonormal to $\e$, i.e., $\E$ lies in the plane perpendicular to $\k$. Due to isotropy, $u^\perp$ alone fully represents the twice degenerated (shear) dynamics.
 By linearizing around the global equilibrium, we write the 
 local Maxwellian contribution to $f$ in (\FINALREF{linear}) as $f^\textrm{LM} = f^\textrm{GM}(1+\varphi_{0})$ 
 where $\varphi_{0}$ takes a simple form, $\varphi_{0} = \X^0\cdot\x$ (linear quasi equilibrium manifold), where $\X^0(\textbf{c})$ was defined in Eq.\ (\FINALREF{X0}). It is conveniently considered as four--dimensional vector using the
 four--dimensional version $\x=[\tilde{n},u^\parallel,\tilde{T},u^\perp]$, and is then given by (\FINALREF{defX0}). It proves convenient to introduce a vector of velocity polynomials, $\mbf{\xi}(\textbf{c})$, which is similar to $\X^{0}$ and defined by (\FINALREF{defXi}), such that $\langle f^\textrm{GM}\xi_{\mu}|f^\textrm{GM}X^0_{\nu}\rangle=\delta_{\mu\nu}$.
 Hence, the fields $\x$ are obtained as $\langle \mbf{\xi}(\textbf{c})\rangle_{f^{LM}}=\x$, where averages are defined as:
 \be 
  \ave{\mbf{\xi}(\textbf{c})}_f = \frac{1}{n_0} \int \mbf{\xi}(\textbf{c}) f(\textbf{c})d^3\textbf{v}
  = \ave{f^\textrm{GM}\mbf{\xi}(\textbf{c})|f}.  \label{defin}
 \ee
 We introduce yet unknown fields $\delta \X(\c,\k)$ 
 which characterize the part $\delta f$ of the distribution function.
 As long as deviations from the local Maxwellian stay small, we seek for 
 a nonequilibrium manifold which is also linear in the hydrodynamic fields $\x$ themselves.
 Therefore, we set: 
 \be \label{expanded} \delta \varphi=\delta \X\cdot \x.\ee
 The ``eigen''-closure (\FINALREF{expanded}), which formally and very generally addresses the
 fact that we wish to {\em not} include other than hydrodynamic
 variables, implies a closure between moments of the distribution
 function, to be worked out in detail below.
 By using the above form (\FINALREF{expanded}) 
 for $\delta f=f^\textrm{GM}\delta\varphi$, with $\hat{L}\delta f = f^\textrm{GM} L[\delta \X]\cdot
 \x$, 
 and the canonical abbreviations $ \triangle \X \equiv\X^{0}(\textbf{c})+ \delta
 \X(\textbf{c},\textbf{k})$, Eq. (\FINALREF{start}) reads: \be
  \omega f^\textrm{GM}\triangle \X \cdot \x
 = \Lambda \Delta f= -i \textbf{k}\cdot\textbf{c} f^\textrm{GM} \triangle \X \cdot \x + f^\textrm{GM}\hat{L}\delta \X\cdot
 \x.
  \label{macf}
 \ee
 The microscopic projected dynamics is obtained from (\FINALREF{IE}) by 
 introducing the thermodynamic projection operator, defined in \cite{karlinbook}, which, when acting upon $J(f)=\Lambda \Delta f$, gives:
 \be P \Lambda \Delta f=D_{\x}\Delta f\cdot\int\mbf{\xi}(\textbf{c})\Lambda \Delta f d^{3}v,\label{proj} \ee
 where $D_{\x}\Delta f\equiv \partial \Delta f/\partial \x$ and the quantity inside the integral in (\FINALREF{proj}) represents the time evolution equations for the moments $\x$. These are readily obtained by integration of the weighted (\FINALREF{start}) as
 \be
  \omega\langle \mbf{\xi}(\textbf{c})\rangle_{f}
   = -i\k\cdot\langle \mbf{\xi}(\textbf{c})\textbf{c}\rangle_{f} + \langle \mbf{\xi}(\textbf{c})\rangle_{\hat{L}\delta f}.
 \label{eocmom}
 \ee
 As shown in Tab.\ \FINALREF{tab:notation} , $D_{\x}\Delta f=f^\textrm{GM}\triangle \X$ holds, whereas (\FINALREF{eocmom}) is linear in $\x$ and can be written as $\omega \x = {\bf M}\cdot \x$. Hence, Eq. (\FINALREF{proj}) attains the form:
  \be
  P \Lambda \Delta f = f^\textrm{GM}\triangle \X\cdot{\bf M}\cdot\x.
 \label{macro} \ee 
 In the derivation of (\FINALREF{macro}), one needs to take into account that $\langle \mbf{\xi}(\textbf{c})\rangle_{\delta f}={\bf 0}$
 (as the fields $\x$ are defined through the local Maxwellian part of the distribution function only) and that $\langle \mbf{\xi}(\textbf{c})\rangle_{\hat{L}\delta f}={\bf 0}$.
 The dependence of the matrix elements of $\bf M$ upon moments of $\delta f$ is explicitly given in Tab.\ \FINALREF{tab1}.
 Combining (\FINALREF{macf}) and (\FINALREF{macro}), and requiring that the
 result holds for any $\x$ (invariance condition), we obtain a
 closed, singular integral equation (invariance equation) for
 complex-valued $\delta \X$,
 \be
   \Delta \X \cdot {\bf M}=  -i \textbf{k}\cdot\textbf{c}\,\triangle \X+\hat{L}\delta \X.
 \label{eqsolve}
 \ee
 Notice that $\delta \X = \Delta \X-\X^0$ vanishes for $k=0$, which implies that 
 the invariant manifold $\Omega_{k\rightarrow 0}$ in that limit is given by the set of local Maxwellians $f^\textrm{LM}$.
 The implicit equation (\FINALREF{eqsolve}) for $\delta \X$ (or $\Delta\X$, as $\X^0$ is known)
 is identical with the
 eigen-closure (\FINALREF{expanded}), and is our main and practically
 useful result. The Bhatnagar-Gross-Krook (BGK) collision model treated in
 \cite{matteo3} is recovered for $\hat{L}(\delta\X)=-\delta \X$.
 
 \subsection{Solving the Invariance Equation} \label{solving}
 
 The invariance equation (\FINALREF{eqsolve}) as well as some symmetry relations
 for the components $\delta X_{\mu}$ of the nonequilibrium
 distribution functions (worked out in Appendix \FINALREF{appendixB} for the interested reader) 
 are exact. Solutions to this equation can be obtained in simple cases. Considering the BGK kinetic equation, for instance, the IE
 could recently be solved numerically and the spectrum of
 hydrodynamic modes at arbitrary wavelength has been successfully
 determined \cite{matteo3}. In the present case, our strategy to
 solve (\FINALREF{eqsolve}) is to confine ourselves with a special kind of interaction potential (Maxwell molecules) and is based upon the results obtained by Chang-Uhlenbeck \cite{uhlenbeck}. They provided an analytical solution to the eigenvalue problem for the BE with the Maxwell molecules collision operator (i.e.: gas molecules interacting via a potential $V\propto
 r^{-4}$, see also \cite{JSP8}). Their analysis showed that due to the isotropy of the operator $\hat{L}$ (i.e. it commutes with rotation operators in velocity space), it admits the following set of eigenfunctions $\Psi_{r,l}({\bf c})$, 
 \be \Psi_{r,l}=\sqrt{\frac{r! (l+\frac{1}{2})\sqrt{\pi}}{
 (l+r+\frac{1}{2})!}}\,c^{l}P_{l}(z)S_{l+\frac{1}{2}}^{(r)}(c^{2}),
 \label{eigfunMax} \ee
 where $P_{l}$ and $S_{l+\frac{1}{2}}$ denote, respectively, Legendre
 and Sonine polynomials, $c=|\c|$ and $z=\c\cdot\e/c$ (see also Appendix \FINALREF{appendixA}). These eigenfuctions are orthonormal with respect to the scalar product (\FINALREF{prod}), with corresponding eigenvalues:
 \be
  \lambda_{r,l} = 2\pi\int \sin(\vartheta) F(\vartheta) T_{r,l}(\vartheta)\,d\vartheta, \label{eigvalMax}
 \ee
 where the explicit expressions for the $F(\vartheta)$ and
 $T_{rl}(\vartheta)$ are needed to numerically solve (\FINALREF{eqsolve}) and hence
 delegated to Appendix \FINALREF{appendixA}.
 Whereas the construction outlined in Sec.\ \FINALREF{secII} deduces the eigenvalues of the perturbed
 system (\FINALREF{eigen}), which vanish in the $k\rightarrow0$ limit,
 just from the knowledge of ker$[\hat{L}]$ (i.e., the ``ground states'' of the unperturbed system), here we attempt a different route. We introduce, first, a decomposition of the microscopic particle velocity, where its components can be expressed through the absolute value of velocity, $c$, and the cosine of the
 angle between velocity and wave vector, denoted as $z$, see (\FINALREF{velocity}). 
 Next, we expand our functions $[\X^{(0)},\delta \X]$
 in terms of the orthonormal basis $\Psi_{r,l}=\Psi_{r,l}(c,z)$:
 \beas{expansion}
 X_{\mu}^{(0)}(c,z)&=&\sum_{r,l}^{N}a_{\mu}^{(0)(r,l)}\Psi_{r,l}(c,z), \\
 \delta X_{\mu}(k,c,z)&=&\sum_{r,l}^{N}a_{\mu}^{(r,l)}(k)\Psi_{r,l}(c,z). 
 \eeas
 The equilibrium coefficients $a_{\mu}^{(0)}$ are known, and can be
 determined, by taking advantage of the orthogonality of the
 eigenfunctions, as:
 \be {\bf a}^{(0)(r,l)} = \pi^{-\frac{3}{2}}\int e^{-c^2}
 \Psi_{r,l}(c,z) \X^{(0)}(c,z)\,d^3{\bf c}. \label{arl0} \ee
 Inserting (\FINALREF{expansion}) into the IE (\FINALREF{eqsolve}), we obtain
 the following nonlinear set of algebraic equations for the unknown
 coefficients $a_{\mu}^{(r,l)}(k)$:
 \bea && \left(a_{\nu}^{(r',l')}+a_{\nu}^{(0)(r',l')}\right)M_{\nu\mu}=\label{IE2}\\
 && -i {\bf k} \cdot \sum_{r,l}^{N}\left(a_{\mu}^{(0)(r,l)}+a_{\mu}^{(r,l)}\right)\mbf{\Omega}_{(r,l,r',l')}+\sum_{r,l}^{N}a_{\mu}^{(r,l)}\mathcal{L}_{(r,l,r',l')} ,
 \nonumber\eea
 with: 
 \beas{L1OM1}
 \mathcal{L}_{(r,l,r',l')}&=&\langle f^\textrm{GM}\Psi_{r',l'}|\hat{L}|f^\textrm{GM}\Psi_{r,l}\rangle, \label{L1} \\
 \mbf{\Omega}_{(r,l,r',l')}&=&\langle f^\textrm{GM}\Psi_{r,l}|\textbf{c}|f^\textrm{GM}\Psi_{r',l'}\rangle. \label{OM1}
 \eeas
 For any order of expansion, the solutions of (\FINALREF{IE2}) characterize an invariant manifold in the phase
 space. The matrix elements $\mathcal{L}_{(r,l,r',l')}$ can be easily evaluated in few kinetic models, as for
 the BGK collision operator, hard spheres and Maxwell molecules. In
 particular, the latter case is recovered by setting:
 \be \mathcal{L}^\textrm{Maxw}_{(r,l,r',l')}=
 \lambda_{r,l}\delta_{r,r'}\delta_{l,l'}. \label{Maxw} \ee
 Furthermore, the simplest case is BGK where
 all nonvanishing eigenvalues attain the constant value:
 $\lambda_\textrm{BGK}=-1$. 
 
 The calculation of the coefficients ${\bf a}^{(r,l)}$ is central in
 our derivation. Through these coefficients, the invariant manifold
 $\Omega\subset U$ is fully characterized: that is, the distribution
 function is determined and the corresponding matrix $\textbf{M}$ of
 linear hydrodynamics is made accessible. Generalized transport coefficients 
 such as viscosity and diffusion coefficients, 
 defined in the Table \FINALREF{tab1}, can be expressed in 
 terms of these coefficients 
 and they enter the definition of the stress tensor and heat flux 
 (explicit expressions provided in Appendix \FINALREF{appendixA}). In the regime of large
 Knudsen numbers the coefficients ${\bf a}^{(r,l)}$ may be further used
 to, e.g., directly calculate phoretic accelerations onto moving and 
 rotating convex particles \cite{phoretic}.

          \INLINEFIGURE{
          \begin{figure*}[tbh]
          \begin{center}
          \includegraphics[width=16.0 cm, angle=0]{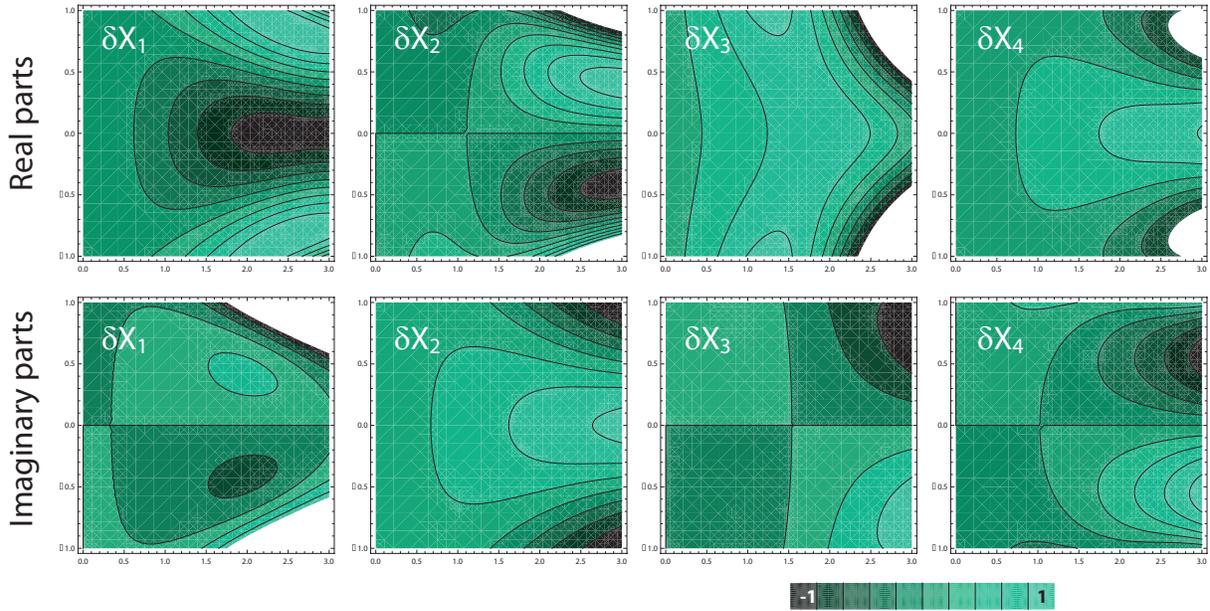}
	  \end{center}
          \caption{{\footnotesize (Color online) All contributions $\delta X_{1-4}({\bf c},{\bf k})$ vs. $c$ (horizontal, $c=|{\bf c}|$) and $z\in[-1,1]$ (vertical axis, $z$ is the cosine of the angle between ${\bf k}$ and peculiar velocity ${\bf c}$) to the nonequilibrium distribution function $\delta f=f^\textrm{GM}\delta X_\mu x_\mu$ (\ref{expanded}) at
$k = 1$, obtained with the fourth order expansion, $N=4$. Shown here are both their real (top) and imaginary parts (bottom row).
}}
          \label{Matteo4_Fig1}
          \end{figure*}   
          }
          \INLINEFIGURE{
          \begin{figure}[tbh]
          \begin{center}
          \includegraphics[width=8.5 cm, angle=0]{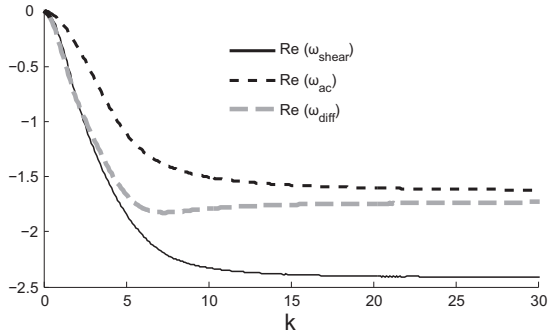}
	  \end{center}
          \caption{{\footnotesize Hydrodynamic modes $\omega$ of the Boltzmann
kinetic equation with Maxwell molecules collision operator as a function of wave number $k$. Shown are two complex conjugated acoustic modes $\omega_\textrm{ac}$, twice degenerated shear mode $\omega_\textrm{sh}$ and a thermal diffusion mode $\omega_\textrm{diff}$.
}}
          \label{Matteo4_Fig2}
          \end{figure}   
          }
 
 \subsection{Hydrodynamic modes and transport coefficients} \label{moresolving}
 
 With ${\bf M}$ at hand, the hydrodynamic modes can finally be obtained from
 (\FINALREF{macro}). The damping rates of the fluctuations (given by the real part of the hydrodynamic modes)
 are obtained by truncating the series (\FINALREF{expansion}) at the 4th order, and represented in
 Fig.~\FINALREF{Matteo4_Fig2}.
 The first important finding is that, for any finite order of
 expansion, the modes extend smoothly over all the wavevector domain
 and, for large $k$, they attain an asymptotic value. This reflects the fact that, below a certain length-scale (more specifically, for lengths less than the mean free path), we reach the free-streaming limit, i.e., the regime in which the collisions cease to occur and particles move along straight lines. Hence, when reducing further the length scale, we may not expect an increase of the damping rate without the ``thermalizing'' effect of collisions. 
 These physical arguments were already supported by the study
 of the BGK kinetic equation \cite{matteo3}, wherein the hydrodynamic
 modes, in the limit of small wavelengths, reach all the same value
 equivalent to the constant eigenvalue of the BGK collision operator.
 A further indication of the role played by the spectrum of
 $\hat{L}$ for large $k$ is provided by the observation that, when taking
 into account all the set of the eigenvalues of $\hat{L}$ which are unbounded
 below,
 also the hydrodynamic modes grow unboundedly.
 
 Generalized transport coefficients are obtained by the nontrivial
 eigenvalues of $-k^{2}\textrm{Re}(\textbf{M})$: $\lambda_{2}=-A$
 (elongation viscosity), $\lambda_{3}=-\frac{2}{3}Y$ (thermal
 diffusivity) and $\lambda_{4}=-D$ (shear viscosity). After some
 algebra it is possible to recast the expression for the higher
 order moments in terms of time correlation functions, in order to
 show the connection with the familiar Green-Kubo expressions. 
 To
 this aim, we first write the nonequilibrium distribution function at
 time $\tau$ as:
 \bea \delta f(\textbf{k},\textbf{c},\tau)&=&e^{\Lambda
 \tau}\delta f(\textbf{k},\textbf{c},0)\label{noneq}. \eea
 
 \begin{table} 
  \begin{tabular}{c@{\quad}c@{\quad}c@{\quad}c}
  \hline\hline
 $\sigma^\parallel_1$ &  $\sigma^\parallel_2$ &$\sigma^\parallel_3$
 &$\sigma_4$\\ 
 $\aveeGMS{\lambda^\parallel \delta X_1}$ &
 $\aveeGMS{\lambda^\parallel  \delta X_2}$ &
 $\aveeGMS{\lambda^\parallel \delta X_3}$ &
 $\aveeGMS{ \gammapara\gammaperpnew \delta Y_4}$ \\
 $-k^2 B$ & $ik A$ & $-k^2 C$ & $ik D$ \\
 real, $\plus$ & imag,$\plus$ & real,$\plus$ & imag,$\minus$ \\
  \hline\hline
 $q^\parallel_1$ &$q^\parallel_2$ &$q^\parallel_3$ &$q_4$ \\
 $\aveeGMS{\gamma^\parallel\delta X_1}$ &
 $\aveeGMS{\gamma^\parallel\delta X_2}$ &
 $\aveeGMS{\gamma^\parallel\delta X_3}$ &
 $\aveeGMS{(\cc^2-\frac{5}{2}) \gammaperpnew\delta Y_4}$ \\
 $ikX$ & $-k^2 Z$ & $ikY$ & $-k^2 U$ 
 \\
 imag,$\minus$ & real,$\minus$ & imag,$\minus$ & real,$\plus$\\
 \hline\hline
  \end{tabular}
        
 \caption{Symmetry adapted components of (nonequilibrium) stress
 tensor $\stress$ and heat flux $\q$, both introduced in (\FINALREF{sq}).
 Row 2: Microscopic expression of these components (averaging with
 the global Maxwellian). Short-hand notation used:
 $\lambda^\parallel=\gammapara^2-\frac{\cc^2}{3}$ and
 $\gamma^\parallel= (\cc^2-\frac{5}{2}) \gammapara $.
 Row 3:
 Expression of the components in terms of (as we show, real-valued)
 functions $A$--$Z$ (see text).
 Row 4: Parity with respect to $z$ -- symmetric ($\plus$) or
 antisymmetric ($\minus$) -- of the part of the corresponding $\delta
 X$ entering the averaging in row 2, and whether this part is
 imaginary or real-valued (see Fig.~\FINALREF{Matteo4_Fig1}).
 Row 3 is an immediate consequence of row 4. }
           \label{tab1}
           \end{table}
 
          \INLINEFIGURE{
          \begin{figure}[tbh]
          \begin{center}
          \includegraphics[width=8.5 cm, angle=0]{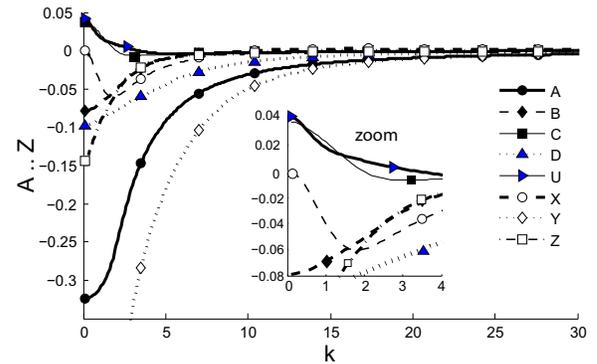}
	  \end{center}
          \caption{{\footnotesize (Color online) Moments $A$-–$Z$ of the distribution function (see Tab.~\ref{tab1} and Eq.~\ref{moments}) 
vs. wave number $k$ obtained with the
solution of (\ref{eqsolve}). Non-triangles (black symbols): Moments entering only the longitudinal component of hydrodynamic equations. Triangles (blue symbols): Moments entering the transverse component of hydrodynamic equations.
}}
          \label{Matteo4_Fig3}
          \end{figure}   
          }
 
 Then, due to (\FINALREF{expansion}), by integrating both sides of (\FINALREF{eqsolve}),
 we find:
 \bea M_{\mu\nu} &=&
 \sum_{r,l}^{N}(a_{\nu}^{(0)(r,l)}\!+\!a_{\nu}^{(r)(r,l)})\langle
 f^\textrm{GM}\xi_{\mu}|\Lambda e^{\Lambda
 \tau}|f^\textrm{GM}\Psi_{r,l}\rangle\nonumber\\
 &&+\sum_{r,l}^{N}a_{\nu}^{(c)(r,l)}\langle
 f^\textrm{GM}\xi_{\mu}|\Lambda e^{\Lambda \tau}|f^\textrm{GM}\Psi_{r,l}\rangle, \label{path} 
 \eea
 where $a_{\nu}^{(r)(r,l)}$ and $a_{\nu}^{(c)(r,l)}$ are, respectively,
 real and imaginary--valued coefficients and $\langle
 f^\textrm{GM}\xi_{\mu}|\Lambda e^{\Lambda \tau}|f^\textrm{GM}\Psi_{r,l}\rangle = -i k\langle
 f^\textrm{GM}\xi_{\mu}|\cpa e^{\Lambda \tau}|f^\textrm{GM}\Psi_{r,l}\rangle$ because lower order moments of the collision operator identically vanish. Next, using the operator identity:	
 \be
 \Lambda e^{\Lambda \tau}= -\Lambda
 \left(\int_{\tau}^{\infty}e^{\Lambda t}dt\right) \Lambda, \label{opid}
 \ee
 we find for the real part of the $\textbf{M}$ matrix in (\FINALREF{path}), for an arbitrary time $\tau=0$:
 \bea \Re(M_{\mu\nu}) &=& -\sum_{r,l}^{N} a_{\nu}^{(c)(r,l)}
 \!\!\int_{0}^{\infty}\!\!dt\langle\dot{\xi}_{\mu}(0)\dot{\Psi}_{r,l}(t)\rangle_{\F},
 \label{GK}\eea
 where $\dot{\Psi}_{r,l}=\Lambda\Psi_{r,l}$. Equation (\FINALREF{GK}) extends to arbitrary wave vector the Green-Kubo relations for transport coefficients. These relations hold in the hydrodynamic regime, when the system, as a result of many collisions, has reached local equilibrium. The opposite regime ($k\gg 1$) is represented by a simple gas of noninteracting point particles. Importantly, as it is evident from Fig.\ \FINALREF{Matteo4_Fig3}, and as already noticed in \cite{alder, forster}, the transport coefficients vanish in the limit of small wavelengths. This is due to the fact that the coefficients $a_{\nu}^{(r)(r,l)}$ and $a_{\nu}^{(c)(r,l)}$, solving (\FINALREF{IE2}), vanish in that limit. This vanishing character of transport coefficients (and, hence, of the heat flux and the stress tensor as is evident from Tab.~\FINALREF{tab1}) for large $k$, corresponds to Eulerian (inviscid) hydrodynamics. We are led, then, to similar conclusions to those traced when we discussed, in Sec. \FINALREF{secIII}, the $k \rightarrow 0$ limit of the invariance equation (\FINALREF{IE}): in the free streaming regime, the local equilibrium manifold (local Maxwellian) becomes an invariant manifold. Let us recall that the Maxwellian distribution constitutes the zero point of the collision integral, in the sense that, in local equilibrium, the net flux of molecules entering and leaving an infinitesimal volume in space, due to the scattering processes, is zero. What we observe here is that, at a sufficiently short length-scale, the distribution function reduces to a Maxwellian, since the contribution from the scattering event, again, vanishes: but now this is because collisions ceased to occur.

 {}
 
          \INLINEFIGURE{
          \begin{figure}[tbh]
          \begin{center}
          \includegraphics[width=8.5 cm, angle=0]{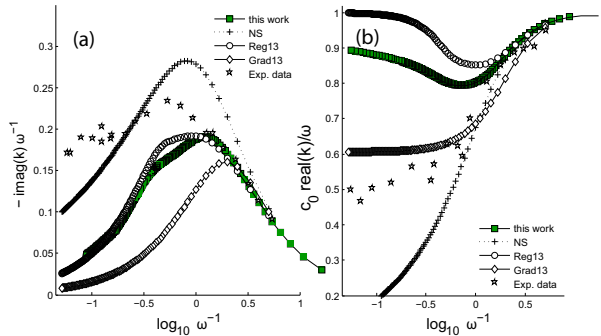}
	  \end{center}
          \caption{{\footnotesize (Color online) (a) Damping spectrum, i.e., the negative imaginary part of $k$ divided by frequency $\omega$ vs. the negative logarithm of $\omega$. Results obtained in this work (by solving Eq.\ \ref{eqsolve}, and subsequently Eq.\ \ref{macro} for $w(k)$ with complex-valued $k$ and real-valued $\omega$) are compared with previous approaches including Navier Stokes (NS), regularized 13 moment (Reg13) \cite{Struchtrup07}, Grad 13 moment (Grad13), and experimental data presented in \cite{exprefmatteo}. (b) Phase spectrum, i.e, real part of $k$ times velocity of sound $c_0$ and divided by $\omega$ vs. the negative logarithm of $\omega$. Again, we compare with reference results.
}}
          \label{Matteo4_Fig4}
          \end{figure}   
          }

 \section{Finite wavelengths hydrodynamics} \label{comparison} \label{secIV}

 We raised the issue of the validity of the notion of invariant manifold for small wavelengths. We were able to show that hydrodynamic modes and the generalized transport coefficients extend smoothly over all the $k$--domain (there is no occurrence of any critical point as in the case of a Grad kinetic system, studied in \cite{CKK12}); hence our approach, here and in \cite{matteo3}, tends to predict that the notion of invariant manifold holds also for short length scales. This would be in agreement with the celebrated papers by Alder {\em et al.} \cite{alder,alder2} who considered a hard spheres gas and showed that hydrodynamic laws remain valid down to times comparable with the time between collisions, $t_\textrm{coll}$, and that the $k$-dependent zero frequency transport coefficients decay until they vanish at short length scales. 
 It would be significant, therefore, to investigate the features of our model at finite frequencies and wavelegths and verify whether the procedure of truncation we introduced in (\FINALREF{expansion}) introduces a length scale below which our coarse-grained description breaks down. 
 In Fig.\ \FINALREF{Matteo4_Fig4} a comparison is shown about inverse phase velocity and damping for acoustic waves between our results, former approaches \cite{grad,Struchtrup07} and experimental data performed by Meyer and Sessler \cite{exprefmatteo}.
 As it is seen, our results are very close to the predictions of the regularized 13 (Reg13) moments method \cite{Struchtrup07}
 and closer to experimental data than Reg13 concerning the phase spectrum. 
 Our theory is capable to predict that the phase speed remains finite also at high frequencies, a feature which is not possessed by any hydrodynamics derived from the CE expansion.
 A further clue about the features of our predictions in the regime of finite frequencies and wavevectors can be achieved by a closer inspection upon the spectrum of density fluctuations.
 To this aim, we introduce the Laplace transform of the hydrodynamic fields $\textbf{x}_{\textbf{k}}(z)=\int_{0}^{\infty}e^{-z t}\textbf{x}_{\textbf{k}}(t)dt$ and write the equation of linear hydrodynamics, analogously to (\FINALREF{macro}), as:    
 \be
 \textbf{x}_{\textbf{k}}(z)=(z \textbf{I}-\textbf{M})^{-1}\textbf{x}_{\textbf{k}}(t=0). \label{laplace}
 \ee     
 By inverting the Laplace transform one obtains:
 \be
 \textbf{x}_{\textbf{k}}(t)=\frac{1}{2 \pi i}\oint\frac{e^{z t}}{(z \textbf{I}-\textbf{M})}dz \,\textbf{x}_{\textbf{k}}(t=0).\label{inverse}
 \ee
 In order to proceed further and calculate the intermediate scattering functions $C_{\textbf{x},\textbf{x}}(\textbf{k},t)=\langle \textbf{x}_{\textbf{k}}(t) \textbf{x}_{-\textbf{k}}(0) \rangle$ it is needed, then, to define the averages which are employed in the calculation of correlation functions. These are, in fact, no longer ensemble averages, as in (\FINALREF{defin}), but, due to (\FINALREF{inverse}), are averages over initial conditions, weighted by the probability density of thermodynamic fluctuation theory \cite{hansen,sengers}. 
 Finally, the power spectrum of $C_{\textbf{x},\textbf{x}}$ is given by its Fourier transform:
 \be
 S_{\textbf{x},\textbf{x}}(k,\omega)=\int_{-\infty}^{\infty}C_{\textbf{x},\textbf{x}}(\textbf{k},t) e^{-i \omega t}dt,
 \ee
 
 It is worth focusing upon the spectrum of density fluctuations, $S_{\tilde{n},\tilde{n}}$, as, being it related to the scattering cross-section, it is a quantity which is experimentally accessible. The calculation of $S_{\tilde{n},\tilde{n}}$ proceeds along the lines indicated above. It just suffices to notice how the solution for $\tilde{n}_{\textbf{k}}(z)$ involves terms proportional to the initial values of $\tilde{n}_{\textbf{k}},\tilde{\u}_{\textbf{k}},\tilde{T}_{\textbf{k}}$, but, following standard recipes \cite{hansen}, only the term proportional to $\tilde{n}_\textbf{k}(t=0)$ needs to be retained in the calculation. By considering just the lower order terms in $k$, one obtains:
 \be
 \tilde{n}_{k}(t)=\left[\frac{2}{5} e^{-\chi k^{2}t}+\frac{3}{10} e^{-\Gamma k^{2} t}\cos (c_{0} k t)\right]\tilde{n}_{k}(0). \label{nkt}
 \ee
 
 The first term in (\FINALREF{nkt}) represents a fluctuation which decays according to a purely diffusive process, with a lifetime proportional to $D_{T}$, whereas the second term represents a fluctuation propagating through the fluid at the (dimensionless) speed of sound $c_{0}=\sqrt{5/3}$ and decaying with a lifetime given by $\Gamma$. The coefficient $D_{T}$ generalizes the standard thermal conductivity, while $\Gamma$ generalizes the combined effect of both thermal conductivity and longitudinal kinetic viscosity. In the limit of small $k$, and following standard text books \cite{reichl}, their expression is given by $D_{T}=\frac {2}{5}(X-Y)$ and $\Gamma=-(\frac {1}{2}A+\frac {1}{5}X+\frac {2}{15}Y)$. Unlike standard treatments of hydrodynamic fluctuations, the generalized transport coefficient $X$ enters the expression of the coefficients $D_{T}$ and $\Gamma$, even though its contribution, as it is evident from Fig.\ \FINALREF{Matteo4_Fig3} is fairly small.
 The (approximate) intermediate correlation function is then obtained by averaging:
 \bea
 C_{\tilde{n},\tilde{n}}(\textbf{k},t)&=&\delta_{\textbf{k},0}+\langle \tilde{n}_{\textbf{k}}(0)\tilde{n}_{-\textbf{k}}(0)\rangle\times 
 \nonumber \\
 && \left[\frac{2}{5} e^{- D_{T} k^{2}|t|}+\frac{3}{10} e^{-\Gamma k^{2}|t|} \cos (c_{0} k t)\right] , \label{corr}
 \eea
 and the dynamical structure factor, hence, attains the following form:
 \bea
 S_{\tilde{n},\tilde{n}}(k,\omega)&=& 
 \delta(\omega)\delta_{k,0}+\langle \tilde{n}_{k}(0)\tilde{n}_{-k}(0)\rangle \times \label{spectrum}\\
 &&\!\!\left[\frac{2}{5}\frac{2 D_{T} k^{2}}{\omega^{2}+(D_{T} k^{2})^{2}}+ \frac{3}{10}\frac{2\Gamma k^{2}}{(\omega \pm c_{0}k)^{2}\!+\!(\Gamma k^{2})^{2}}\right]. \nonumber
 \eea
 
          \INLINEFIGURE{
          \begin{figure}[tbh]
          \begin{center}
          \includegraphics[width=8.5 cm, angle=0]{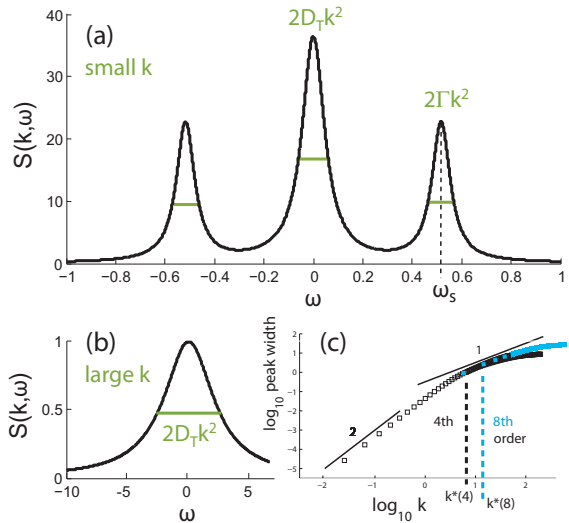}
	  \end{center}
          \caption{{\footnotesize (a) Dynamic structure factor $S(k,\omega)$ vs. $\omega$ for a small $k=0.4$ and (b) large $k=100$. $\omega_s=c_{0}k$ denotes the hydrodnamics predicted sound mode of the spectrum, and the widths are related to the moments $A$--$Z$ (see Fig.~~\ref{Matteo4_Fig3}). For small $k$, these are given by $D_T = \frac{2}{5}(X-Y)$ and $\Gamma=-(\frac {1}{2}A+\frac {1}{5}X+\frac {2}{15}Y)$, where $A$ is the generalized longitudinal kinetic viscosity, $Y$ the generalized thermal diffusion coefficient and $X$ is a cross-coupling transport coefficient, relating heat flux to density gradients. (c) Width $D_Tk^2$ of the Rayleigh peak vs. $k$ (double-logarithmic). At small $k$, $D_Tk^2\propto k^2$ as all moments $A$--$Z$, except $X$, reach a finite value in this limit. The inflection point at $k=k^*(N)\gg 1$ (shown to be increasing with the order of expansion $N$) denotes the onset of departure from the ideal Maxwellian behavior, where the width of the peak starts to behave sublinearly in $k$, and is used to quantify the range of validity for results obtained at finite order.
}}
          \label{Matteo4_Fig5}
          \end{figure}   
          }
 
 Representative plots of $S(k,\omega)$ are shown in Figs.~\FINALREF{Matteo4_Fig5}a--b. 
 For small $k$ (hydrodynamic limit), the spectrum we obtain recovers the usual results of neutron (or light) scattering experiments and consists of three Lorentzian peaks. The one centered in $\omega=0$ is the Rayleigh peak, which corresponds to the diffusive thermal mode. The two side peaks centered in $\omega\pm c_{0}k$ are the Brillouin peaks, and represent the two propagating sound waves.
 By increasing the wave-vector, the structure of (\FINALREF{spectrum}) is unchanged except that the generalized coefficients $D_{T}$ and $\Gamma$ need to be replaced by more complicate expressions, not given here. The net effect observed is that sound waves get strongly damped and vanish, whereas the central Rayleigh peak decreases and broadens. Density fluctuations are, therefore, driven only by a diffusive thermal mode for large enough $k$.
 A deeper look about the behavior of the width at half maximum of the central Rayleigh peak with increasing wavevectors 
 allows us to bridge the gap between the hydrodynamic continuum-like description and the free particle limit.
 The hydrodynamic regime is featured by a width increasing with the square wavevector, $\propto k^{2}$. On the contrary, in the free particle limit, the calculation of the dynamical structure factor $S_{n,n}(k,\omega)$ reduces to the Fourier transform of the self part of the van Hove function $G_{s}(\textbf{r},t)$\cite{hansen}, which, upon writing $\textbf{c}=\textbf{r}/t$, is given by the Maxwellian distribution: $G_{s}(\textbf{r},t)=\pi^{-3/2}v_{T}^{-3}t^{-3}\exp(-r^{2}/t^{2})$. Hence, the width of the peak is expected to grow up linearly in $k$, for large $k$.
 Our results, see Fig.~\FINALREF{Matteo4_Fig5}c, predict a width which is truly quadratic for small enough $k$, reach the 
 regime of linear behavior and terminate, for some large $k$, with a sub-linear dependence on $k$. The onset of the  terminal regime at $k=k^*(N)$ marks the range of validity which can be accessed at a given finite order of expansion, $N$. 
 Increasing $N$ thus does not alter the overall picture we obtained at a moderate order of expansion, 
 and more generally, results obtained with $N+1$ will not change those obtained with $N$ below $k^*(N)$, cf.
 Fig.~\FINALREF{Matteo4_Fig5}c.

 \section{Conclusions}  \label{secV}
 The main result of our paper is the characterization of the nonequilibrium distribution function, through the method of 
 invariant manifolds, and the calculation of its moments (the functions $A$--$Z$), which constitute the building blocks of the generalized hydrodynamic equations.
 As we had previously shown in \cite{matteo2}, the latter equations are stable and hyperbolic for arbitrary wavevectors.
 Moreover, we have proposed and applied a route to solve the eigenvalue problem associated with the BE (\FINALREF{start}), 
 by calculating the hydrodynamic modes, which we may regard either as decay rates of hydrodynamic 
 fluctuations as well as generalized eigenfrequencies of the BE (\FINALREF{BElin}).
 The generalized transport coefficients have been numerically determined and settled into expressions 
 recovering the Green-Kubo formulas. Finally, also by comparing with available experimental data and previous approaches, we discussed the range of validity of our approach, which turned out to be capable of extending the hydrodynamic scenario to length scales below the mean free path. This offers new perspectives towards a deeper comprehension of the transition between a 
 ``mesoscopic'' particle-like description of matter and the ``continuum'' macroscopic one.
 
 \subsubsection*{Acknowledgement}
 
 M.C. thanks Dr. I.V. Karlin for helpful discussions. 
 This work was supported by EU-NSF contract NMP3-CT-2005-
 016375 and FP6-2004-NMP-TI-4 STRP 033339 of the European
 Community

 \begin{appendix}
 
 \section{Obtaining the invariance equation}\label{appendixA}
 
 In order to calculate the averages occurring in Sec.\ \FINALREF{secIII}, 
 like $\langle\mu(\textbf{c})\rangle_{f}$, we switch to spherical coordinates.
 For each (at present arbitrary) wave vector $\k=k\e$, we choose the coordinate system in such a way that its
 (vertical) $z$-direction aligns with $\textbf{e}_\parallel$ and that its 
 $x$--direction aligns with $\textbf{e}_\perp$. The velocity vector 
 we had been decomposed earlier as $\tilde{\u}=u^\parallel\e + u^\perp \E$. 
 We can then express
 $\textbf{c}$, over which we are going to perform all integrals, 
 in terms of its norm $c$, a vertical variable $z$ and
 plane vector $\textbf{e}_{\phi}$ (azimuthal angle
 $\textbf{e}_{\phi}\cdot\textbf{e}_\perp=\cos\phi$; the plane contains $\E$) for the present
 purpose as:
 \be 
 \textbf{c}/c = \sqrt{1-z^2}\,\textbf{e}_{\phi} + z
 \textbf{e}_\parallel, \label{velocity}
 \ee 
 as shown in Fig.~\FINALREF{Matteo4_Fig6}.
 
          \INLINEFIGURE{
          \begin{figure}[tbh]
          \begin{center}
          \includegraphics[width=8.5 cm, angle=0]{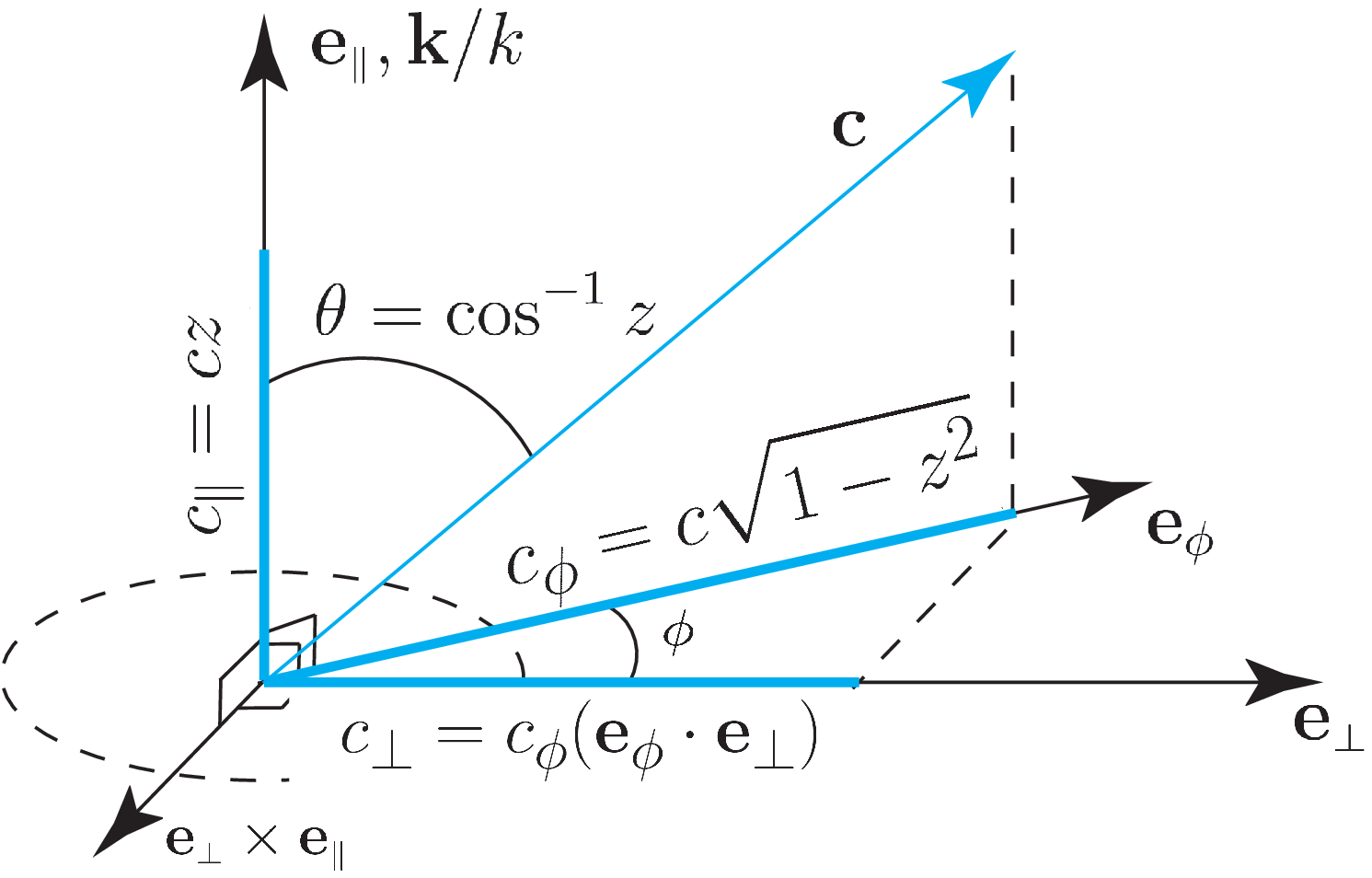}
	  \end{center}
          \caption{{\footnotesize Schematic drawing introducing a orthonormal frame $\e$, $\E$, and $\E\times\e$ which is defined by the wavevector $\k \,\|\,\e$ and the heat flux $\q$ (not shown), which lies in the $\e$--$\E$--plane. Shown is the velocity vector $\c$ (\ref{velocity}) relative to this frame (characterized by length $c$, coordinate $z$, and angle $\phi$) and its various components. The integration over $d^3c=c^2dcdzd\phi$ is done in spherical coordinates with respect to the local orthonormal basis.
}}
          \label{Matteo4_Fig6}
          \end{figure}   
          }
 
 The local Maxwellian, linearized around global equilibrium, takes the form:
 $f^\textrm{LM}/f^\textrm{GM}=1+\varphi_0= 1+\X^0\cdot\x$,
 where the four--dimensional $\X^0$, and the related vector $\mbf{\xi}$, employing four--dimensional $\x=[\tilde{n},u^\parallel,\tilde{T},u^\perp]$, 
 are given by
 \beas{defsXXi}
  \X^0(\textbf{c}) &=& \left(1,2 \cpa,(c^2-\frac{3}{2}),2 \cphi\right), \label{defX0} \\
  \mbf{\xi}(\textbf{c})&=&\left(1,\cpa,\frac{2}{3}(\cc^2-\frac{3}{2}),\cphi\right). \label{defXi}
 \eeas Here, we
 introduced, for later use, the abbreviations
 \be
  \cpa \equiv \textbf{c}\cdot\textbf{e}_\parallel,
 \qquad 
 \cphi \equiv \textbf{c}\cdot\textbf{e}_\perp,
 \qquad 
  \cpe \equiv \c\cdot\textbf{e}_\phi = \frac{\cphi}{\textbf{e}_\perp \cdot\textbf{e}_{\phi}},
 \label{defc}
 \ee
 such that $i\k\cdot \textbf{c}=ik \cpa$. We can then rewrite
 (\FINALREF{velocity}) as 
 $\textbf{c}=\cpe\textbf{e}_{\phi} + \cpa \textbf{e}_\parallel$
 with $\cpa=cz$ and $\cpe=c\sqrt{1-z^2}$. The latter two components, 
 contrasted by $\cphi$ (and $\textbf{e}_{\phi}$), do not depend on the azimuthal
 angle. We further introduced yet unknown fields $\delta \X(\c,\k)$
 which characterize the nonequilibrium part of the distribution
 function, $\delta\varphi = \delta f/f^\textrm{GM}$. By analogy with the structure of the local Maxwellian, those are linear in terms of the hydrodynamic fields $\x$ themselves,
 \be \label{expanded2} \delta \varphi=\delta \X\cdot \x =
 \delta X_1\tilde{n} +\delta X_2u^\parallel + \delta X_3\tilde{T} + \delta X_4
 u^\perp. \ee
  The functions $\delta X_{1,2,3}$, which are associated to the longitudinal fields, inherit the full rotational symmetry of the corresponding Maxwellian components, $\delta X_{1,2,3}=\delta X_{1,2,3}(c,z)$, whereas
 $\delta X_4$ factorizes as $\delta X_4(c,z,\phi) = 2\delta Y_4(c,z)
 \sum_{m=1}^{\infty} y_m \cos m\phi$. 
 In this context it is an important technical aspect of our derivation to work with a suitable orthogonal set of
 basis functions (irreducible tensors, cf. \cite{irr}, for models beyond the Maxwell gas) to represent $\delta f$
 uniquely. 
 The matrix ${\bf M}$ in (\FINALREF{macro}) contains the
 non-hydrodynamic fields, the heat flux $\q \equiv
 \avee{\c(\cc^2-\frac{5}{2})}_{f}$ and the stress tensor
 $\mbf{\sigma}\equiv \avee{\overline{\c\c}}_{f}$, where $\bar{\bf s}$
 denotes the symmetric traceless part of a tensor ${\bf s}$
 \cite{CKK12,tensorrechnung2,mkbook}, $\overline{\bf s}=\frac{1}{2}({\bf s}+{\bf
 s}^T) - \frac{1}{3}{\rm tr}({\bf s})\ONE$. 
 Using (\FINALREF{expanded}) and the above mentioned angular dependence of the $\delta \X$ functions (the only term in $\delta X_4$ playing a role in our calculations is the first order term $\cos\phi$, with $y_1=1$, see \cite{calc}), 
 constraints, such as the required decoupling between longitudinal and transversal dynamics of the hydrodynamic fields, 
 are automatically dealt with correctly when performing integrals over $\phi$.
 More explicitely \cite{calc}, the stress tensor and heat flux uniquely decompose as follows
 \beas{sq}
  \mbf{\sigma} &=& \sigma^\parallel\,\frac{3}{2}\overline{\e\e} + \sigma^\perp\,2\overline{\e\E}, \\
  \q &=& q^\parallel\,\e + q^\perp\,\E, 
 \eeas
 with the moments $\sigma^\parallel =
 (\sigma^\parallel_1,\sigma^\parallel_2,\sigma^\parallel_3)\cdot(\tilde{n},u^\parallel,\tilde{T})$
 and $\sigma^\perp = \sigma_4u^\perp$, and similarly for $\q$ (see Row 2
 of Tab.~\FINALREF{tab1}). The prefactors arise from the identities
 $\overline{\e\e}:\overline{\e\e}=\frac{2}{3}$ and
 $\overline{\e\E}:\overline{\e\E}=\frac{1}{2}$.
 We note in passing that, while the stress tensor has, in general,
 three different eigenvalues, in the present symmetry adapted
 coordinate system it exhibits a vanishing first normal stress
 difference.
 Since the integral kernels of all moments in (\FINALREF{sq}) do not
 depend on the azimuthal angle, these are actually two-dimensional
 integrals over $c\in[0,\infty]$ and $z\in[-1,1]$, each weighted by a
 component of 
 $2\pi \cc^2 f^\textrm{GM}\delta \X$. Stress tensor and heat flux can
 yet be written in an alternative form which is defined by Row 3 of
 Tab.~\FINALREF{tab1}. As we will prove below, due to fundamental symmetry
 considerations, the hereby introduced generalized transport coefficients $A$--$Z$ are
 real-valued. They can be expressed in terms of the moments of the distribution
 function, i.e., expansion coefficients ${\bf a}^{(r,l)}$,  
 as follows:
 \bea A&=&-\frac{i a_{2}^{(0,2)}}{\sqrt{3} k},
 \quad B=-\frac{ a_{1}^{(0,2)}}{\sqrt{3} k^{2}},\nonumber\\
 C&=&-\frac{a_{3}^{(0,2)}}{\sqrt{3} k^{2}},
 \quad X=-\frac{i \sqrt{5} a_{1}^{(1,1)}}{2 k},\nonumber\\
 Y&=&-\frac{i \sqrt{5} a_{3}^{(1,1)}}{2 k}, \quad Z=-\frac{\sqrt{5} a_{2}^{(1,1)}}{2 k^{2}},\nonumber\\
 D&=&-\frac{i}{k}\sum_{r,l}^{N}a_{4}^{(r,l)}\langle f^\textrm{GM}\cpa \cpe|f^\textrm{GM}\Psi_{r,l}\rangle,\nonumber\\
 U&=&-\frac{1}{k^{2}}\sum_{r,l}^{N}a_{4}^{(r,l)}\langle
 f^\textrm{GM}(\cc^2-\frac{5}{2}) \gammaperpnew|f^\textrm{GM}\Psi_{r,l}\rangle.
 \label{moments}
  \eea
 
 We proceed by using these functions
 $A$--$Z$ to split ${\bf M}$ into parts as ${\bf M} = \Re({\bf M}) -
 i \,\Im({\bf M})$,
 \be
  {\bf M} = k^2 \left(\begin{array}{cccc}
  0 & 0 & 0 & 0 \\
 0 & A & 0 & 0 \\
 \frac{2}{3} X & 0 & \frac{2}{3}Y & 0 \\
 0 & 0 & 0 & D
 \end{array}\right)
 -i k \left(\begin{array}{cccc}
  0 & 1 & 0 & 0 \\
 \tilde{B} & 0 & \tilde{C} & 0 \\
 0 & \tilde{Z} & 0 & 0 \\
 0 & 0 & 0 & 0
 \end{array}\right), \label{checkerboard}
 \ee
 with abbreviations $\tilde{B}\equiv\frac{1}{2}\!-\!k^2 B$,
 $\tilde{C}\equiv\frac{1}{2}\!-\!k^2 C$, and
 $\tilde{Z}\equiv\frac{2}{3}(1\!-\!k^2 Z)$.
 The checkerboard structure of the matrix ${\bf M}$
 (\FINALREF{checkerboard}) is particularly useful for studying properties
 of the hydrodynamic equations (\FINALREF{macro}), such as hyperbolicity
 and stability (see \cite{CKK12} and below), once the functions
 $A$--$Z$ are explicitly evaluated. We remind the reader that we use orthogonal basis functions
 (irreducible moments, cf. Tab.~\FINALREF{tab1}) to solve (\FINALREF{eqsolve}).
 In order to show how the above functions enter the definition of the ${\bf M}$ matrix, we first
 notice that its elements are -- a priori -- complex valued. We wish,
 then, to make use of the fact that all integrals over $z$ vanish for
 odd integrands. To this end we introduce abbreviations $\plus$
 ($\minus$) for a real-valued quantity which is even (odd) with
 respect to the transformation $z\rightarrow -z$. One notices
 $\X^0=(\plus,\minus,\plus,\plus)$, and we recall that $A$--$Z$ are
 integrals over either even or odd functions in $z$, times a
 component of $\delta \X$ (see Tab.~\FINALREF{tab1}).
 
 Let us prove the consistency of the specified symmetry of {\bf M}
 and the invariance condition: Start by assuming $A$--$Z$ to be
 real-valued functions. Then $M_{\mu\nu}=\plus$ if $\mu+\nu$ is even,
 and $M_{\mu\nu}=i\plus$ otherwise. This implies $\delta
 X_1=\plus+i\minus$, $\delta X_2=\minus+i\plus$, $\delta
 X_3=\plus+i\minus$, and $\delta X_4=\plus+i\minus$, i.e., different
 symmetry properties for real and imaginary parts. With these
 ``symmetry'' expressions for $\X^0$, $\delta \X$, and ${\bf M}$ at
 hand, 
 and by noticing that symmetry properties for 
 $\delta \X$ take over to $\hat{L}(\delta \X)$ because the $\psi_{r,l}$ 
 are (i) symmetric (antisymmetric) in $z$ for even (odd) $l$ and 
 (ii) eigenfunctions of $\hat{L}$, 
 we can insert into the right hand side of the equation,
 $\hat{L}(\delta \X) = (\X^0+\delta \X)\cdot({\bf M}+i\minus{\bf
 I})$, which is identical with the invariance equation
 (\FINALREF{eqsolve}).
 There are only two cases to consider, because ${\bf M}$ has a
 checkerboard structure, i.e., only two types of columns:
 Columns $\mu=1$ and $\mu=3$:
 $\delta X_\mu
 = \plus + i \minus$ because $M_{1-3,4}=0$;
 Columns $\mu\in\{2,4\}$: $\delta X_\mu
 = \plus + i \minus$ if $M_{\mu,1-3}=0$ (which is the case for column
 $4$) and $\minus+i\plus$ if $M_{\mu,4}=0$ (which is the case for
 column $2$). These observations complete the proof.

 \section{Exact solution to the
 eigenvalue problem for a Maxwell-molecules collision operator}
 \label{appendixB}
 
 Given the linearized Boltzmann collision operator:
 \bea L\delta f&=& \int\int d\Omega d \textbf{c}_{1} \sigma(\Omega,g)
 g f^\textrm{GM}(c_{1}) \times \nonumber \\
 && [\dphi(\textbf{k},\textbf{c})\!+\!\dphi(\textbf{k},\textbf{c}_1) \!-\! \dphi(\textbf{k},\textbf{c}')\!-\!\dphi(\textbf{k},\textbf{c}_{1}']\eea
 where $g=|v-v_{1}|$ is the absolute value of the relative velocity
 and $\sigma(\Omega,g)$ the differential collision cross section. 
 For so-called Maxwell molecules the collision probability per unit time
 is independent of the relative velocity:
 \be g \sigma(\Omega,g)=\sqrt{\frac{2 K (M+m)}{M m}}F(\vartheta), \ee
 where $m$, $M$ are the masses of the colliding particles and
 $F(\vartheta)$, with $\vartheta\in[0,\pi]$, is given in parametric form
 through the parameter $\phi\in[0,\pi]$: \bea
  \vartheta(\phi) &=& \pi - 2\sqrt{\cos(2\phi)} K(\sin\phi), \label{paratheta} \\
  F(\vartheta)    &=& \frac{[2^{3/2}\sin\vartheta \sin(2\phi)]^{-1}
    \sqrt{\cos 2\phi}}{\cos^2(\phi) K(\sin\phi) -\cos(2\phi) E(\sin\phi)} \label{paraF} ,
 \eea 
 with the elliptic integrals 
 $K(x) = \int_0^{\pi/2} (1-x^2\sin^2 y)^{-1/2}\,dy$, and $E(x) = \int_0^{\pi/2} (1-x^2 \sin^2
 y)^{1/2}\,dy$. 
 Since the collision operator is spherically symmetric
 in the velocity space, the dependence of the eigenfunctions upon the
 direction of $\textbf{c}$ is expected to be spherically
 harmonic. Indeed, the eigenvalue problem admits the following solutions:
  \bea
  L[\psi_{r,l}(c,z)] &=& \lambda_{r,l} \psi_{r,l}(c,z), \\
  \psi_{r,l}(c,z)   &=& \sqrt{\frac{r!(l+\frac{1}{2})\sqrt{\pi}}{(l+r+\frac{1}{2})!}} \;c^l P_l(z) S_{l+\frac{1}{2}}^{(r)}(c^2),
  \label{eigenfunctions}
 \eea where $S_{l+1/2}^{(r)}(x)$ are Sonine polynomials, and $P_l(z)$
 are Legendre polynomials  which act on the azimuthal component of
 the peculiar velocity ${\bf c}$. The Legendre and Sonine polynomials
 are each orthogonal sets, \bea
  \int_{-1}^1 P_l(z) P_{n}(z) \,dz &=& \frac{2}{2l+1} \delta_{ln}, \nonumber \\
  2\pi \int_0^\infty c^2 e^{-c^2} c^{2l} S_{l+\frac{1}{2}}^{(r)}(c^2) S_{l+\frac{1}{2}}^{(p)}(c^2) \,dc
   &=&  \frac{\pi (l+\frac{1}{2}+r)!}{r!}\delta_{rp}. \nonumber
 \eea Accordingly, the $\psi_{r,l}$ are normalized to unity with the
 weight factor $f_0(c)=\pi^{-3/2}\exp(-c^2)$ (as defined in Sec.\ \FINALREF{2A}): \bea
  \delta_{rr'}\delta_{ll'} &=&
   2\pi^{-1/2} \int_{-1}^1\!\int_0^\infty  c^2 e^{-c^2} \psi_{r,l}(c,z)\psi_{r',l'}(c,z)\, dc dz \nonumber \\
  &\equiv& \pi^{-3/2}\int e^{-c^2} \psi_{r,l}({\bf c})\psi_{r',l'}({\bf c}) \,d^3c.
 \label{ortho} \eea The corresponding eigenvalues for Maxwell molecules are given by:
 \beas{eigenvalues}
  \lambda_{r,l} &=& 2\pi \int \sin(\vartheta) F(\vartheta) T_{rl}(\vartheta)\,d\vartheta , \\
    T_{rl}(\vartheta) &\equiv& \cos^{2r+l}\left(\frac{\vartheta}{2}\right) P_l(\cos\frac{\vartheta}{2})
           +\sin^{2r+l}\left(\frac{\vartheta}{2}\right) P_l(\sin\frac{\vartheta}{2}) \nonumber \\
          &&  - (1+\delta_{r0}\delta_{l0}),
 \eeas
 
 The collision operator is negative semidefinite, that is, all eigenvalues are
 negative except $\lambda_{0,0}$, $\lambda_{0,1}$, and $\lambda_{1,0}$ which are zero and correspond to the collision invariants.
 As it was shown in \cite{uhlenbeck}, there is no lower bound
 for the set of eigenvalues. 
 Chang and Uhlenbeck's investigation \cite{uhlenbeck} of the dispersion of sound in a Maxwell molecules gas was based upon writing the deviation from the global equilibrium as:
 $(\varphi_{0}+\delta\varphi)=\sum_{\{r,l\}=0}^{\infty}a_{r,l}\Psi_{r,l}(\textbf{c})$
 so that the eigenvalue equation reduces to an algebraic equation for
 the coefficients $a_{r,l}$:
 \bea
  \omega a_{r,l}&=&-i \textbf{k}\cdot \!\!\!\sum_{\{r',l'\}=0}^{\infty}\!\!\mbf{\Omega}_{r,l,r',l'}a_{r',l'}+
  \lambda_{r,l}a_{r,l}, \label{Chang} \\
 \mbf{\Omega}_{r,l,r',l'}&\equiv&\langle f^\textrm{GM}\Psi_{r,l}|\textbf{c}|f^\textrm{GM}\Psi_{r',l'}\rangle.
 \eea
 
 The hydrodynamic modes for the Maxwell-molecules gas are found by setting to zero the determinant of the above system
 of linear equations. Within this approach, from the knowledge of the
 spectrum of $\hat{L}$, it is possible to solve the eigenvalue
 problem (\FINALREF{BElin}) for an arbitrary number of modes, just by
 tuning the number of eigenfunctions taken into account in the
 \textit{ansatz} for the nonequilibrium distribution function. The
 peculiarity of the Maxwell-molecules gas lies in the fact that at
 any stage of approximation the modes recover and extend those
 corresponding to lower order approximations. 
 This method produces results which are found to be in agreement with the CE
 expansion.

 \end{appendix}

\end{document}